\documentclass{elsarticle}
\usepackage{lipsum}
\makeatletter
\def\ps@pprintTitle{%
 \let\@oddhead\@empty
 \let\@evenhead\@empty
 \def\@oddfoot{}%
 \let\@evenfoot\@oddfoot}
\makeatother

\usepackage[utf8]{inputenc}

\usepackage{float}

\usepackage{listings}
\usepackage{color}

\definecolor{codegreen}{rgb}{0,0.6,0}
\definecolor{codegray}{rgb}{0.5,0.5,0.5}
\definecolor{codepurple}{rgb}{0.58,0,0.82}
\definecolor{backcolour}{rgb}{0.95,0.95,0.92}

\definecolor{identifiercolor}{rgb}{.8,.6,.56}
\definecolor{inactivecolor}{rgb}{0.15,0.15,0.5}
 
\lstdefinestyle{mystyle}{
    backgroundcolor=\color{backcolour},   
    commentstyle=\color{codegreen},
    keywordstyle=\color{magenta},
    numberstyle=\tiny\color{codegray},
    stringstyle=\color{codepurple},
    identifierstyle={\bfseries\color{identifiercolor}},
    basicstyle=\tiny, 
    breakatwhitespace=false,         
    breaklines=true,                 
    captionpos=b,                    
    keepspaces=true,
    showspaces=false,                
    showstringspaces=false,
    showtabs=false,                  
    tabsize=2,
    language=Mathematica
}
 
\usepackage{url}

\usepackage{graphicx}
\usepackage{float}
\usepackage[caption = false]{subfig}

\usepackage{natbib}

\begin{document}

\title{Computer simulations combined with experiments 
for calculus-based physics laboratory course}


\author[sergey]{Sergey V. Samsonau}
\ead{ssamsonau@gradcenter.cuny.edu}
\address[sergey]{Princeton International School of Mathematics and Science, Princeton, NJ, USA}

\begin{abstract}
This paper is presenting a set of laboratory classes to be taught as a part of a 1-year calculus-based physics class. It is composed out of 7 modules designed to bring together experiments and computer simulations. Each module uses both simulations and experiments to address a phenomenon under study, and lasts for 3 weeks (21 weeks total for the whole set). Wolfram Mathematica is used for computer simulations. Topics are: Motion with a drag, Pendulum with an arbitrary amplitude, Magnus effect, Centripetal force acting on a pendulum, A ball rolling from paths of various shapes, Doppler shift and Fourier transformation, Equipotential lines. These laboratory classes were taught for 2 years at Princeton International School of Mathematics and Science.
\end{abstract}

\maketitle

\section{Introduction}
In the recent decades computer simulations became an important way to do research in many areas. Complexity of many real system does not allow to obtain a model written through analytic equations. In such situations computer simulations provide a good (if not the only one) way to do address them. Computer simulations allow people to perform weather forecasting \cite{Miller2008}, understand dynamics of star clusters \cite{Aarseth1973}, enhance particle physics research \cite{Karsch1993}, allow to perform simulations of chemical interactions \cite{Gissinger2017}, as well as address many other systems. Modern research often relies on combination of simulations and actual experiments to get more information about the same phenomenon \cite{samsonau2013, Ozone2009}. It is our believe that simulations should be introduced to physics students as early as possible, and this should be done in a combination with actual experiments in the lab. 

Goging back in history, we can find, that the question about educating people in the area of simulations was raised as early as in 1965 \cite{Egbert}. However, up to now simulations did not become a part of a standard curriculum neither for the 1st or 2nd year of physics curriculum of a typical US physics program \cite{Halliday2014, Giancoli2014, CollegeBoard2014}. A question of how to include simulations into a curriculum and how efficient such inclusion is remains to be open \cite{Rutten2012, Bradley2014, Sarabando2014, Bozkurt2010}. To understand the issue deeper, we would like to highlight the distinct ways computational physics education is approached. 

\begin{itemize}
    \item Large variety of specially designed computational physics classes exist that are taught after students take regular physics classes  \cite{Mazvovsky2012, Burke2016, Landau2007, Martin2016, Caballero2017, MATLAB2007}. In these classes students do not perform any actual experiments with equipment. Thus students just have to "believe" that simulations actually are a good way to describe reality.  
    \item Linking physics to computer games in regular physics classes \cite{Bourg2004, Price2006} serves as a great way to motivate students. However, this approach still asks students to "believe" that simulations work, just because they are based on an appropriate theoretical model.
    \item Use of simulations for virtual experiments \cite{Bradley2014, PhET2006}. While being a great way to extend (or replace) demonstrations and experiments, this approach does not aim to teach students to do by themselves neither actual experiments or simulations.
\end{itemize}

We believe that introducing computer simulations in a combination with experiments allows students to see and feel a direct connection between those, as well as understand their limitations. In such settings a design of simulations and experiments go hand in hand, and should be developed to be consistent with each other. Doing that inside a standard physics course allows to make sure that all students taking physics get exposure to at least 3 ways one can do science: theory, experiment, and computer simulations. This gives students a better picture on what kind of work is possible in science, and may positively influence one's decision to continue education in physics. 

An inclusion of simulations into a standard physics course should be done with a consideration to time limitations: in order to add something, something shall be removed or modified. Our thought is that a modification of a laboratory part of a course may be a good way to introduce simulations combined with experiments: students will still work on improving their experimental skills and at the same time will learn simulations. In this case instead of doing many small labs, students will do lower number of longer labs. This is also a great opportunity for many instructors to convert their labs from typical step-by-step labs into open-inquiry experiments. 

Another consideration should be given to the fact that many students come to physics course without any previous programming experience, and thus there is not enough time to teach them coding in Python (see the reference \cite{Bourg2013} for physics simulations with Python) or other language like that. Teaching programming would also shift a course focus from physics, which would be counter productive. Wolfram Mathematica language provides an easy to understand, follow (and modify) way to perform good computer simulations (with default settings) without a need of teaching students a full scale programming and numerical methods class. Just an installation itself can be a deal-breaker, as Python tends to require to pay a lot of attention to package dependencies (especially taking into account various operation systems students can use), while Wolfram Mathematica works without any issues right after an installation. As a side benefit, students exposed to Mathematica language discover for themselves a powerful way to perform symbolic computations. While Mathematica has a great and extensive help with many interactive examples, some teachers and students may find useful to use books introducing Wolfram Mathematica \cite{Wolfram, Hastings}. 
 
To the best of our knowledge there is no published set of laboratory classes for calculus-based physics course available designed to teach students to perform computer simulations together with real experiments. 

This set of labs is designed in a paradigm of open inquiry labs, where there is no one definite answer, and students essentially perform a small research projects. Many decisions have to be made to reach a final result, such as setting up a model, designing an experiment, choosing/making equipment for an experiment, determining a way to evaluate errors, and also importantly the way to compare results of simulation to results of real experiments. Results students obtain vary significantly depending on the decisions students make. Benefits of open-inquiry (and research-oriented) teaching approach are described elsewhere \cite{Murray2016, Mohlhenrich2018, Chinn2002}. These labs were developed with a limited time for lab work in mind, giving students enough time to read about new concepts and implement code and experiments. The code provided bellow can be made smaller (in some cases), but it was delivered in this way for a better readability by students.

 The set was taught for two years in AP physics class in the following settings: 40 minutes once a week of class time (plus time allocated by students for work outside of classroom), 3 weeks per module. It is important to note, that most of the students who were doing these labs took algebra-based physics class with an extensive open-inquiry laboratory classes, and were taught to handle open ended tasks, design experiments, analyze errors, and take responsibility for their own results. Completed work is submitted by students in a “lab article” format (instead of “lab report”), which is constructed to be as close as possible to a typical research paper. The details about “lab article” format will be presented and discussed elsewhere.

\section{Description of Laboratories}

In many of the modules students are responsible for designing their own experiments and are allowed to use all available equipment in the physics laboratory. As these assignments have a research component, students are in fact encouraged to develop their own devices in addition to equipment available in physics laboratory. In case there is a strong engineering program, they can benefit from resources available there as well. Of course, students can ask for an assistance from an instructor, but the instructor should not do the work for students and give them enough freedom and challenge to explore and learn.

Thus in summary students do:
\begin{itemize}
  \item Develop their own experiments to address a specified phenomenon
  \item Develop (or adjust) computer simulations
  \item Compare the results of experiments with the ones produced by computer simulations
  \item Redesign experiments or simulations if needed
  \item Report their results
\end{itemize}

Files with a code presented bellow can be downloaded from this link\footnote{\url{https://github.com/ssamsonau/NumSim_Exp_PhysLabs}} in the Wolfram Mathematica notebook format. Please note that in most cases the presented code does not solve a problem completely and serves as a "starter code", which provides a sufficient base from which students (and instructors) can develop their own scripts. Thus, an instructor have to be willing to learn and practice Wolfram Mathematica language in order to be able to guide and assist students (and in order to be able to evaluate their work). In a case we would present here a code solving  problems completely, we would remove a large portion of a "research aspect" of these laboratory classes. In practice, students tend to come up with an original and interesting code, which can vary greatly in a way a problem was approached. 

\subsection{Motion with a drag}
\subsubsection{Task with potential subtasks}

Study motion with a drag:	
\begin{itemize}
    \item Model free fall motion with a drag
    \item Determine which drag type to use (turbulent vs laminar flow)
    \item Make corresponding plots
    \item Design and perform experiment, to verify model. Experiment can contain modifications to address separate issues under study
\end{itemize}

\subsubsection{Distribution of work by days} Work is distributed by days as follows:

\textbf{Day one.}

\begin{itemize}
    \item An instructor makes an intro to Mathematica: basic operations, plotting, solving equations, etc. 
    \item Students are asked to describe free fall with a kinematic equation and plot y(t) using an analytical equation. 
    \item After that the instructor gives a basic example of use of NDSolve with free fall in no-air-resistance situation. 
    \item Next step is to compare NDSolve solution to kinematic equation. 
    \item Last step is to compare to an experiment (drop a heavy ball). 
    \item Students are asked to learn (at home) how to use NDSolve to solve differential equations. They shall find and read general theory related to Euler and Runge-Kutta methods.
\end{itemize}

\textit{Homework:} At home students will make Mathematica program to integrate an equation of motion with a drag. They should (read and) decide which drag to use: laminar or turbulent. Students also design an experiment in which they can measure an effect of a drag. For example, one can use water in a high and wide graduated cylinder (we use 2000mL Polymethylpentene cylinder from FLINN scientific). Then students can either drop a small object to sink there (note: dense objects will do this too fast), or produce bubbles to observe how they rise (bubbles can be produced by putting a foam under water)

\textbf{Day two.}
Instructor will verify that students have a working code for situation with a drag. If they do, they can do perform experiment.

\textit{Homework:} At home students analyze experimental data and think whether the designed and performed experiment allows to observe effects of a drag. If it is not the case, they should think about another design. Another task is to enter into Mathematica program parameters of a real experiment and compare results between experiment and simulation. 

\textbf{Day three.}
After students compare results between computer simulation and real experiment at home they may need to redesign experiment and redo it. This day provides opportunity to do that.

\subsubsection{Code: starter code} 
\begin{figure}[H]
    \centerline{%
        \resizebox{1\textwidth}{!}{\includegraphics{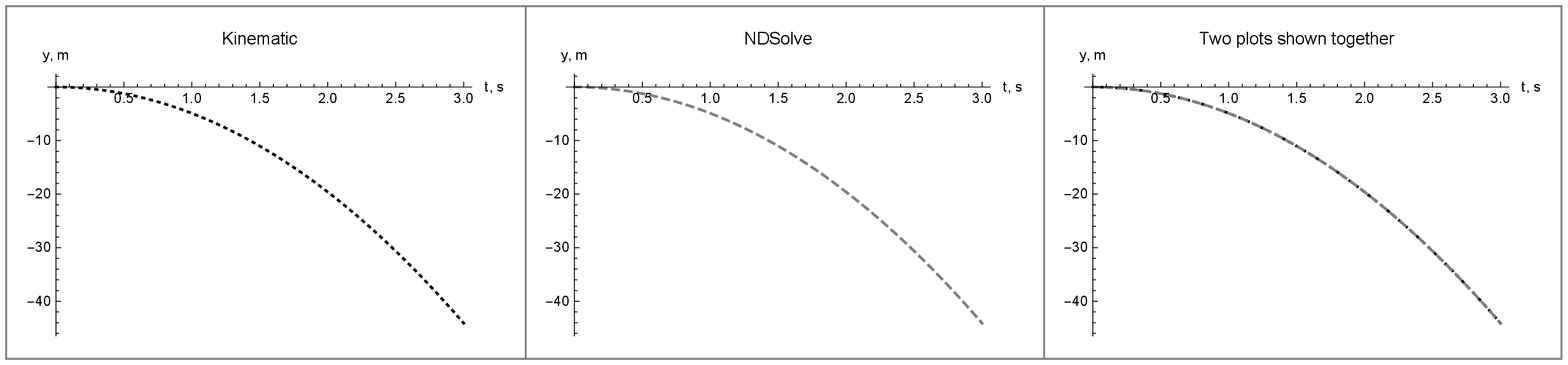}}%
    }
    \caption{Comparison of coordinate vs time dependence for a falling object predicted by a kinematic equation (left) and step-by-step computer simulation (center). The plot of the right shows both predictions plotted together.} 
    \label{fig:kin_step_compare}
\end{figure}

The code bellow is not using any new physics concepts. After execution it will produce the plot shown on fig. \ref{fig:kin_step_compare}

\begin{lstlisting}[caption={Set up path to save files in a current directory},captionpos=t]
SetDirectory[NotebookDirectory[]];
\end{lstlisting}

\begin{lstlisting}[caption={Use a kinematic equation},captionpos=t]
g = - 9.8;
yKin[t_, y0_, v0_]:= y0 + v0 t + g t^2/2;
Plot[yKin[t, 0, 0], {t, 0, 2}, AxesLabel -> {"t, s", "y, m"}];
\end{lstlisting}

\begin{lstlisting}[caption={Use a differential equation},captionpos=t]
s = NDSolve[{y''[t]==g, y[0]==0, y'[0]==0}, y, {t, 0, 5}];
Plot[Evaluate[y[t]/.s], {t, 0, 3}, PlotRange -> All, AxesLabel -> {"t, s", "y, m"}]
\end{lstlisting}

\begin{lstlisting}[caption={Show 3 plots together},captionpos=t]
p1 = Plot[yKin[t, 0, 0], {t, 0, 3}, PlotStyle -> {Black, Dotted},  
          AxesLabel -> {"t, s", "y, m"}, ImageSize -> 300, PlotLabel -> "Kinematic"];

p2 = Plot[Evaluate[y[t]/.s],{t,0,3}, PlotStyle -> {Gray, Dashed},  
          AxesLabel -> {"t, s", "y, m"}, ImageSize -> 300, PlotLabel -> "NDSolve"];

p3 = Show[p1, p2, PlotLabel -> "Two plots shown together"];
p4 = Deploy@Grid[{{p1, p2, p3}}, Dividers -> Gray, Spacings -> {2, 2}]
Export["p4.eps", p4];
\end{lstlisting}

\subsubsection{Code: add drag}

\begin{figure}[H] 
    \centerline{%
        \resizebox{1\textwidth}{!}{\includegraphics{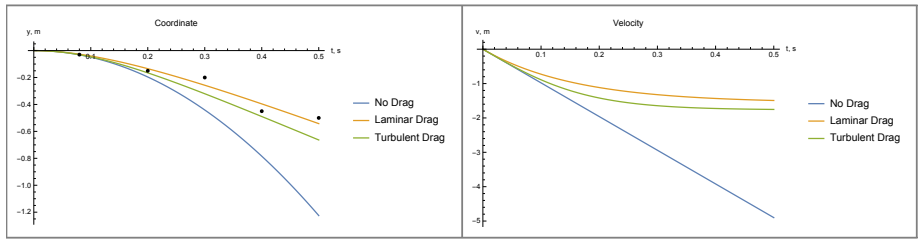}}%
    }
    \caption{Compare models adjusted for drag. Left plot showing y coordinate vs time for a falling object, right plot - a velocity of the said object. "No drag" is drawn in blue, laminar in yellow, turbulent in green. Dots represent sample data to be produced by students.} 
    \label{fig:drag}
\end{figure}

The code bellow have to be adjusted by students. In order to do this, they have to find information and read about models physics uses to describe laminar flow, turbulent flow, and what is Reynolds number. After execution code will produce the plot shown on fig. \ref{fig:drag}

\begin{lstlisting}[caption={Including drag into the model. Will produce a plot shown on fig. \ref{fig:drag}},captionpos=t]
k = 1;  (* used in LaminarDrag bellow *) (* students have to adjust this *)
rho = 1; (* students have to adjust this *); 
Cd = 1; (* students have to adjust this *)
A = Pi (0.01) 2; (*cross section area*) (* students have to adjust this *)
m = 0.01; 

LaminarDrag[v_]:= -Sign[v] k Abs[v] A
TurbulentDrag[v_]:= -Sign[v] 1/2 rho v^2 Cd A

Fg = m g;
Tmax = 0.5;

(* Solve *)
sFree = NDSolve[{y''[t]==Fg /m, y[0]==0, y'[0]==0}, y, {t,0,Tmax}];
sL = NDSolve[{y''[t]==(Fg + LaminarDrag[y'[t]])/m, y[0]==0, y'[0]==0}, y, {t,0,Tmax}];
sT = NDSolve[{y''[t]==(Fg + TurbulentDrag[y'[t]])/m, y[0]==0, y'[0]==0}, y, {t,0,Tmax}];

(*import actual data*)
data = Import["sample.csv"]; (*use //Table to see it as a table*)
pdata = ListPlot[data, PlotStyle -> Black];

(* Plot *)
pY = Plot[{Evaluate[y[t]/.sFree], Evaluate[y[t]/.sL], Evaluate[y[t]/.sT]}, {t, 0, Tmax}, 
      PlotRange -> All, PlotLegends -> {"y No Drag", "y Laminar", "y Turbulent"}, 
      AxesLabel -> {"t, s", "y, m"}, PlotLabel -> "Coordinate"]

pV = Plot[{Evaluate[y'[t]/.sFree], Evaluate[y'[t]/.sL], Evaluate[y'[t]/.sT]}, {t, 0, Tmax}, 
      PlotRange -> All, PlotLegends -> {"v No Drag", "v Laminar", "v Turbulent"}, 
      AxesLabel -> {"t, s", "y, m"}, PlotLabel -> "Velocity"]

pp = Deploy@Grid[{{pY, pV}}, Dividers -> Gray, Spacings -> {2, 2}]
Export["pp.eps",pp];

\end{lstlisting}

\subsubsection{Code: error evaluation and analysis}

\begin{lstlisting}[caption={Error evaluation: possibly useful commands. Will produce a plot shown on fig. \ref{fig:error_evaluation}},captionpos=t]
dataTime = data[[All, 1]]; (* time values in data *)
dataCoordinate = data[[All, 2]]; (* coordinate values in data *)

Prediction = Evaluate[y[ dataTime ]/.sL] [[1]]; (* corresponding prediction from simulation *)

(* Standart Deviation of a difference *)
StDev = StandardDeviation[Prediction - dataCoordinate]; 

(* Maximum Percentage Error *)
MaxPercentageError = Max[100*(Prediction - dataCoordinate)/(0.5*(Prediction + dataCoordinate))];

(* add error bar to plot *)
Needs["ErrorBarPlots`"] (* load package *)
InstError = 0.04; (* set up an instrumental error *)
error = ConstantArray[InstError, Length[data]]; (* add instrumental error *) 
(* students can extend this approach to include accidental error *)

withError=Transpose[{data[[All, 1]], data[[All, 2]], error}];
pdataNew = Show[ListPlot[data], ErrorListPlot[withError, PlotStyle -> Black]];

(* Plot *)
pYError = Show[Plot[{Evaluate[y[t]/.sFree], Evaluate[y[t]/.sL], Evaluate[y[t]/.sT]}, {t, 0, Tmax},
                      PlotRange -> All, PlotLegends -> {"No Drag", "Laminar Drag", "Turbulent Drag"}, 
                      AxesLabel -> {"t, s", "y, m"}, ImageSize -> 400, PlotLabel -> "Coordinate"], 
                      pdataNew];


(* separate plot for errors *)
erp1 = Plot[{InstError}, {x, 0, Tmax}, PlotStyle -> Gray]; 
erp2 = Plot[{-InstError}, {x, 0, Tmax}, PlotStyle -> Gray];

pError = ListPlot[Transpose[{dataTime,  dataCoordinate - Prediction}], 
                  PlotLabel -> "Error", AxesLabel -> {"t, s", "Prediction - Actual, m"},
                  ImageSize -> 400, PlotStyle -> Black];
pErrorT = Show[pError, erp1, erp2];
ppError = Deploy@Grid[{{pYError, pErrorT}}, Dividers -> Gray, Spacings -> {2, 2}]

Export["ppError.eps",ppError];
\end{lstlisting}

\begin{figure}[H]
    \centerline{%
        \resizebox{1\textwidth}{!}{\includegraphics{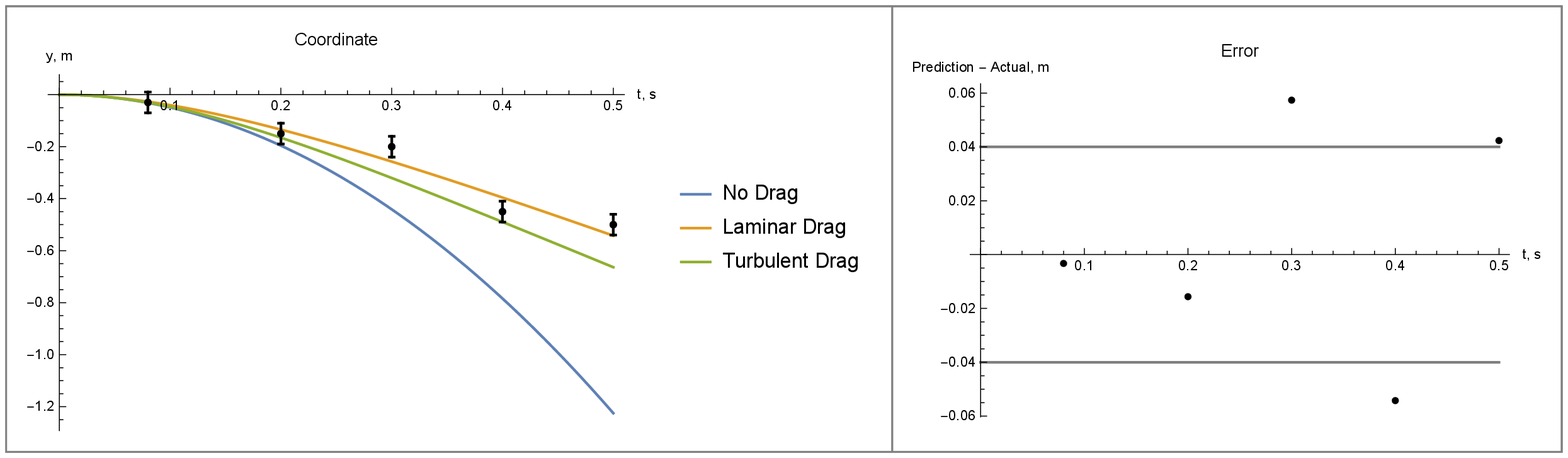}}%
    }
    \caption{Error evaluation example. Coordinate of a falling object vs time on the left. "No drag" is drawn in blue, laminar in yellow, turbulent in green. Dots with error-bars (instrumental error in this case) represent sample data to be produced by students. A graph for a difference between prediction from a model and collected data is displayed on the right. Gray horizontal lines represent error limits.}
    \label{fig:error_evaluation}
\end{figure}

\subsection{Pendulum with an arbitrary amplitude}

\subsubsection{Task with potential subtasks}

Study motion of a real pendulum. The simple formula for simple pendulum is valid in an approximation of small amplitude, but what will happen if the angle is larger?	
\begin{itemize}
    \item Model motion with any amplitude
    \item Study influence of mass
    \item Study influence of drag
    \item Make corresponding plots
    \item Make animation of a motion
    \item Design and perform experiment, to verify model. Experiment can contain modifications to address separate issues under study
\end{itemize}

\subsubsection{Distribution of work by days}

Students are free to schedule their work on their own. An instructor will provide starter code on the first and second days.

\textbf{Day one.} Starter code for pendulum with any amplitude is provided. Using this code students can start developing a model, think how to obtain a period, etc. Optional: students can find a period automatically (by finding maximum of phi(t)) and do a plot of period vs amplitude.
Also an instructor gives a code to generalize drag force using Piecewise function of Mathematica. Some students may choose to add drag force to their equations. 

\textbf{Day two.} An instructor helps to write a code to create animations. Students have to adjust this code on their own to take drag into account. The instructor shall not give the code to produce this result, allowing students to do this on their own.

\textbf{Day three.} Students are free to choose how to use this time.

\subsubsection{Code: starter code}

The code bellow will model a motion of a pendulum.
\begin{lstlisting}[caption={Calculate a period of a simple pendulum with small amplitude},captionpos=t]
SetDirectory[NotebookDirectory[]];

L = 0.5; 
g = -9.81;
T = 2 * Pi * Sqrt[L/Abs[g]]
\end{lstlisting}

\begin{lstlisting}[caption={Describe motion of a pendulum with any amplitude using computer simulation. Will produce a plot shown on fig. \ref{fig:pendulum_angle}},captionpos=t]
m = 0.1;

Ft[phi_]:= -m Abs[g] Sin[phi]

MomInertia = m * L2

s = NDSolve[{phi''[t]==(Ft[phi[t]] * L)/MomInertia, phi[0]==Pi/2, phi'[0]==0}, phi, {t, 0, 10}];

p1 = Plot[{Evaluate[phi[t]/.s], Evaluate[phi'[t]/.s], Evaluate[phi''[t]/.s]}, {t, 0, 10}, 
        PlotRange -> All, PlotLegends -> {"phi", "phi'", "phi''"}]
         
Export["pendulum_p1.eps", p1]         
\end{lstlisting}

\begin{figure}[H]
    \centerline{%
        \resizebox{1\textwidth}{!}{\includegraphics{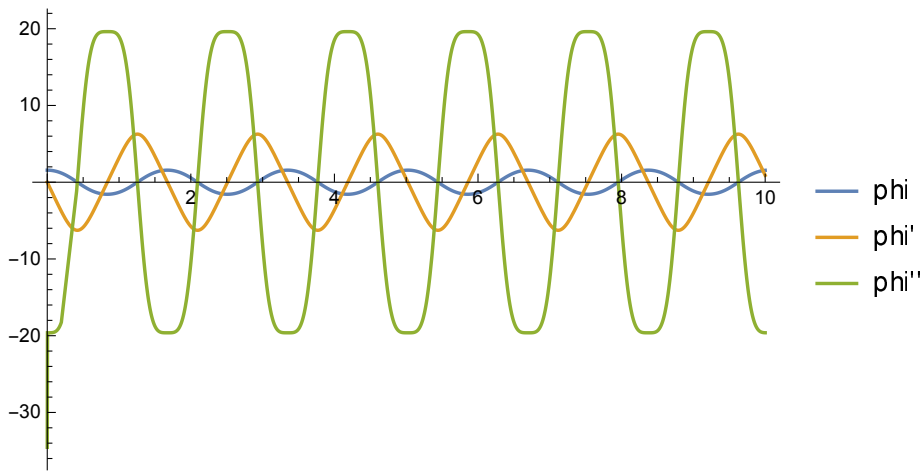}}%
    }
    \caption{Angle in degrees (blue) between a pendulum and the vertical axis, angular velocity (yellow), angular acceleration (green) vs time.}
    \label{fig:pendulum_angle}
\end{figure}

\begin{lstlisting}[caption={Compare period at different amplitudes. Will produce a plot shown on fig. \ref{fig:pendulum_large_amplitude}},captionpos=t]
s2 = NDSolve[{phi''[t]==(Ft[phi[t]] * L)/MomInertia, phi[0]==Pi/2, phi'[0]==0}, phi, {t, 0, 10}];
s6 = NDSolve[{phi''[t]==(Ft[phi[t]] * L)/MomInertia, phi[0]==Pi/6, phi'[0]==0}, phi, {t, 0, 10}];
s10 = NDSolve[{phi''[t]==(Ft[phi[t]] * L)/MomInertia, phi[0]==Pi/10, phi'[0]==0}, phi, {t, 0, 10}];

p2 = Plot[{Evaluate[phi[t]/.s2], Evaluate[phi[t]/.s6], Evaluate[phi[t]/.s10]}, {t, 0, 10},
           PlotRange -> All, PlotLegends -> {"Pi/2", "Pi/6", "Pi/10"}]
Export["pendulum_p2.eps",p2]
\end{lstlisting}

\begin{figure}[H]
    \centerline{%
        \resizebox{1\textwidth}{!}{\includegraphics{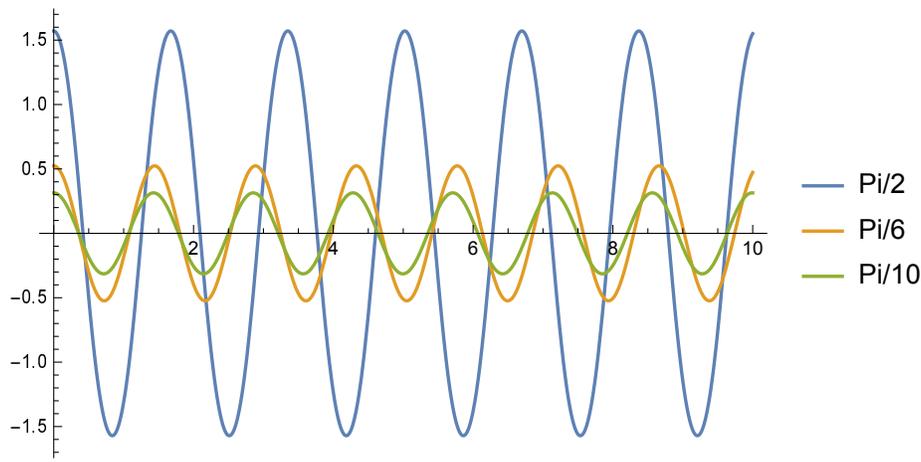}}%
    }
    \caption{Angle (in radians) vs time for different amplitudes: Pi/2 (blue), Pi/6 (yellow), Pi/10 (green).}
    \label{fig:pendulum_large_amplitude}
\end{figure}

\subsubsection{Code: animation, period}

Bellow one can find a code to produce an animation of a moving pendulum.
 
\begin{lstlisting}[caption={Animation of a pendulum, using harmonic dependence},captionpos=t]
x[t_]:= L*Sin[phi[t]]
y[t_]:=-L*Cos[phi[t]]

MaxTime = 2; 

AnF := Animate[
  Graphics[
   {Red, PointSize -> Large, Point[{x[t], y[t]}/.sol], 
    Green, Thick, Line[{{0,0}, {x[t], y[t]}}/.sol]},
    PlotRange -> {{-1.1 * L, 1.1 * L}, {-1.1 * L, 0}}, Axes -> True, Ticks -> False],
  {t, 0, MaxTime, MaxTime/100}, AnimationRunning -> False]
AnF/.sol -> s (* Animation *)

\end{lstlisting}

\begin{lstlisting}[caption={Animation of a pendulum using results of computer simulation},captionpos=t]
AnF/.sol -> s10 (* Animation *)
\end{lstlisting}

\subsection{Magnus effect}

\subsubsection{Task with potential subtasks}

Study motion affected by Magnus effect. In this case the motion should be described using differential equations for two coordinates (x and y). 	
\begin{itemize}
    \item Model motion in 2 dimensions which is not affected by the Magnus effect. Students have to remember that drag force direction will change, according to the change in velocity direction.
    \item Model motion affected by the Magnus effect. Students have to remember that direction of both Drag and Magnus forces will change, according to the change in velocity direction.
    \item Study influence of object’s radius (for example for a cylindrical object)
    \item Study influence of a rotation speed
    \item Make corresponding plots
    \item Design and perform experiment, to test a computational model. Students can choose to address other parameters of the experiment as they see fit.
\end{itemize}

\subsubsection{Distribution of work by days}

Students are free to schedule their work on their own. A starter code can be shown as an example on a screen, but not given to students (allowing students to write their own implementation). Assistance can be given to those who needs help. \textit{Note:} Students will design their own experiments. For an initial demonstration of an effect an instructor can use a paper cylinder (from a sheet of paper) rolling from an incline. Different radii cylinders can be used. Alternatively, two paper cups connected on the bottom with a tape, can be launched with a rubber band, as can be found on some of YouTube videos.

\subsubsection{Code: Magnus effect} Students have to modify the code bellow to match the system they use in an actual experiment.

\begin{lstlisting}[caption={Code to model a motion of a falling object in 2 dimensions (with and without drag, Magnus effect not included)},captionpos=t,
label={lst:2d}]
SetDirectory[NotebookDirectory[]];
rho = 1.2; (*kg/m^3, air density*)
R = 0.0125; (* radius *)
Cd = 1.17; (* drag coefficient *)
TurbulentDrag[v_]:= -(1/2) rho v^2 Cd A ;(* if turbulent drag is used *)

A = Pi R^2; (*crossection area*)
m = 0.01 ;(*10g*)
g = -9.8; 

V[vx_, vy_]:= Sqrt[vx2 + vy2];
CosAlpha[vx_, vy_]:= vx/V[vx, vy];
SinAlpha[vx_, vy_]:= vy/V[vx, vy];

Tmax = 1.5;

sFree = NDSolve[{m * x''[t]==0, m * y''[t]==m g,
    x[0]==0, x'[0]==5, y[0]==0, y'[0]== 0}, {x, y}, {t, 0, Tmax}];

sT = NDSolve[{
    m * x''[t]==TurbulentDrag[V[x'[t], y'[t]]] * CosAlpha[x'[t], y'[t]],
    m * y''[t]==m g + TurbulentDrag[V[x'[t], y'[t]]] * SinAlpha[x'[t], y'[t]],
    x[0]==0, x'[0]==5, y[0]==0, y'[0]==0}, {x, y}, {t, 0, Tmax}];

pF = ParametricPlot[
   Evaluate[{x[t], y[t]}/.sFree], {t, 0, Tmax}, AxesLabel -> {"x coordinate, m", "y coordinate, m"}, 
   PlotLabel -> "Trajectory of motion in 2D with drag", AspectRatio -> Automatic, PlotStyle -> Blue];

pD = ParametricPlot[
   Evaluate[{x[t], y[t]}/.sT], {t, 0, Tmax}, AxesLabel -> {"x coordinate, m", "y coordinate, m"},
   PlotLabel -> "Trajectory of motion in 2D", AspectRatio -> Automatic, PlotStyle -> {Dashed, Red}];
(* Show[pF, pD]; *)

\end{lstlisting}

\begin{lstlisting}[caption={This code will add the Magnus effect to the model created above (using cylinder as an object)},captionpos=t, 
label={lst:magnus}]
L = 0.2; (*cylinder length*)
r = 0.02; (*cylinder radius *)
omega=10; (*assume it is a constant during motion, 
            for a first approximation *)

(*Students have carefully consider sign of omega *)
G = 2 Pi r 2 omega;
Magnus[v_ ]:= L rho v G

sM = NDSolve[{
    m * x''[t]==TurbulentDrag[V[x'[t], y'[t]]] * CosAlpha[x'[t], y'[t]] +
                Magnus[V[x'[t], y'[t]]] * CosAlpha[x'[t], y'[t]],
    m * y''[t]==m g + TurbulentDrag[V[x'[t], y'[t]]] * SinAlpha[x'[t], y'[t]] - 
                Magnus[V[x'[t], y'[t]]] * SinAlpha[x'[t], y'[t]],
    x[0] == 0, x'[0]==5, y[0]==0, y'[0]==0}, {x, y}, {t, 0, Tmax}];

pM = ParametricPlot[Evaluate[{x[t], y[t]}/.sM], {t, 0, Tmax}, 
                    AxesLabel -> {"x coordinate, m", "y coordinate, m"},
                    PlotLabel -> "Trajectory of motion in 2D, and Magnus effect", 
                    AspectRatio -> Automatic, PlotStyle -> Dotted];

pT = Legended[Show[pF, pD, pM], LineLegend[{Blue, {Dashed, Red}, {Dotted, Green}},
                                           {"Free", "Drag", "Magnus"}]]
Export["Mangus_plot.eps", pT];
\end{lstlisting}

\begin{figure}[H]
    \centering
    \includegraphics[width=7cm,height=7cm,keepaspectratio]{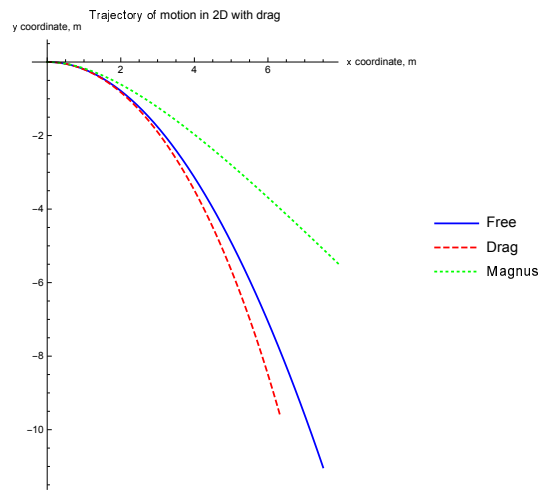}%
    \caption{Trajectory of a falling object in 2 dimensions (y vs x). Free motion (no drag) in blue, motion with drag in red, motion with drag and the Magnus effect in green. Plot created by a code in listings \ref{lst:2d} and  \ref{lst:magnus}}
\end{figure}
\subsection{Centripetal force and Pendulum}

\subsubsection{Task with potential subtasks}

Study a force acting on a string suspension point produced due to a pendulum motion. In this lab students will compare results from an experiment, a computer simulation, and from an analytical expression.

\begin{itemize}
    \item Make a numerical model
    \item Study the influence of various factors
    \item Obtain an analytical expression for the force acting on a suspension point (first, assuming no air drag, and then with air drag)
    \item Make corresponding plots
    \item Design and perform experiment, to verify model. Experiment can contain modifications to address separate issues under study
    \item Compare results produced by three approaches: analytical, numerical, experimental
\end{itemize}

\subsubsection{Distribution of work by days}

Students are free to schedule their work on their own. No starter code is given this time.

For experimental part students can use any force sensor they wish. In our lab we use a force sensor from Vernier company, with a data interface to Vernier Logger Pro software. Students can import collected data into Mathematica using ‘Import’ command and plot it on the same graph with points produced by simulation. Routing of a thread holding the bob of a pendulum from the force sensor should be done in such a way, that there is only longitudinal pull on a force sensor.

\subsubsection{Code}
Students write a code on their own. No new starter code needed for this lab.
\subsection{A ball rolling from paths of various shapes.}

\subsubsection{Task with potential subtasks}

Given tracks of irregular shapes, students will create a numerical model of a motion of a ball rolling along that tracks. This enforces an idea of use of computer simulation to describe a motion of an object given conditions we can’t describe analytically. For example, motion of a space ship moving near planets, large asteroids, and other objects.

\begin{itemize}
    \item Make images of tracks, and obtain coordinates describing their shapes
    \item Perform an experiment
    \item Make a numerical model of a ball rolling down the tracks
    \item Make corresponding plots
    \item Compare results produced by computer simulation with experimental data
\end{itemize}

\subsubsection{Distribution of work by days}

Students are free to schedule their work on their own. An instructor will provide starter code on the first day. 

We use PSworks Conservation of Energy Track from Flinn Scientific, but any other setup with several different paths can be used. Ball rolls down each path with different dependence of coordinate and velocity vs time. Students can use a video camera in a smartphone to obtain experimental dependence of those values and then compare them with a prediction of a model.

In a simplified approach, only a time of the whole motion can be used to compare prediction of model to observation.

In order to make a numerical model, students will have to extract x-y coordinates combination describing the shape of each track. This can be done by making a photo of a track, importing it into Mathematica, and then using a “Coordinates tool” available under the image displayed in Mathematica to extract coordinates. Subsequent interpolation is required to obtain a function, derivative of which will be calculated. Moving average smoothing can be used before an interpolation. Alternatively, a video processing tools of Mathematica can be used to extract the information about trajectory and a dependence of coordinates vs time automatically (the code for that part is provided below as well). 

\subsubsection{Code: starter code}

\begin{figure}[H]
    \centerline{%
        \resizebox{1\textwidth}{!}{\includegraphics{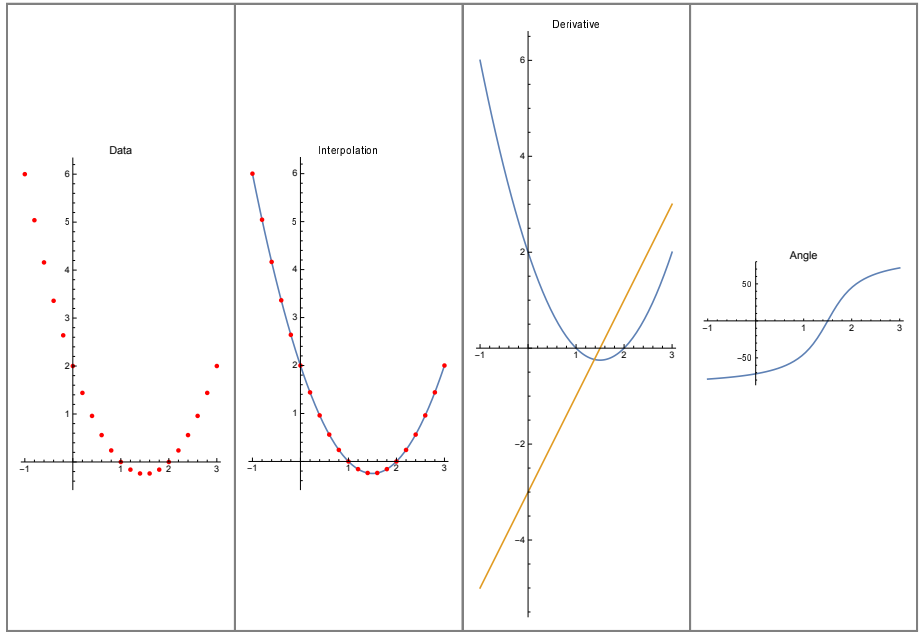}}%
    }
    \caption{Example of a processing of data for a curve. From left to right: plot of dots (which may be obtained from a photo of a track using the Mathematica's "coordinate tool") describing a shape of a track, interpolation, derivative (blue - original curve, yellow - derivative), angle.}
    \label{fig:interpolation}
\end{figure}

As mentioned above, students may obtain a list of points from a photo of a track using the "Coordinates tool". Let us generate points to be used as an example and do a typical processing steps to extract an angle dependence from the x coordinate. When students know a dependence of angle vs x, they can use it to write a motion equation to be used in NDSolve. The summary of processing is presented in fig. \ref{fig:interpolation}

\begin{lstlisting}[caption={Generating dots to be used for an example bellow},captionpos=t]
(*generating a list of point to be used as an example *)
a = Table[{i, i^2 - 3*i + 2 + 0.0 * Random[]}, {i, -1, 3, 0.2}];
ListPlot[a,  AspectRatio -> Automatic, PlotStyle -> {PointSize[Medium], Red}]
\end{lstlisting}

\begin{lstlisting}[caption={Interpolation},captionpos=t]
b = Interpolation[a]; (** b now is a function of x**)
Show[Plot[b[x], {x, -1, 3}], ListPlot[a, PlotStyle -> {PointSize[Medium], Red}], 
    AspectRatio -> Automatic]
\end{lstlisting}

\begin{lstlisting}[caption={Find an angle of the curve},captionpos=t]
Der = b'; (** derivative, or tan(alpha) **)
Plot[{b[x], Der[x]}, {x, -1, 3}, AspectRatio -> Automatic]
angle [x_]:= ArcTan[Der[x]]/Degree;
Plot[angle[x], {x, -1, 3}]
\end{lstlisting}

\subsubsection{Code: automatic Image processing}

As an alternative students can use an automatic image processing available in Mathematica. For that a use of a static, preferably monotone background is a advised to produce a video. Students may choose to use a slow motion option of their smartphones to have a sufficient time resolution. After that the video file can be imported into Mathematica.

\begin{lstlisting}[caption={Import a video file (.mov or .mp4)},captionpos=t]
SetDirectory[NotebookDirectory[]];
p = Import["video.MOV", "ImageList"];
(*p = Import["video.mp4", {"AVI","ImageList"}];*)
Length[p]
\end{lstlisting}

\begin{lstlisting}[caption={Display selected frames. Output is shown on fig. \ref{fig:ball_on_track}},captionpos=t]
p1 = Show[p[[1]] , ImageSize -> 300];
p2 = Show[p[[10]] , ImageSize -> 300];
p3 = Show[p[[27]] , ImageSize -> 300];
pp = Deploy@Grid[{{p1, p2, p3}}, Dividers -> Gray, Spacings -> {2, 2}]
SetDirectory[NotebookDirectory[]];
Export["rolling_down_frames.eps", pp];
\end{lstlisting}

\begin{figure}[H]
    \centerline{%
        \resizebox{1\textwidth}{!}{\includegraphics{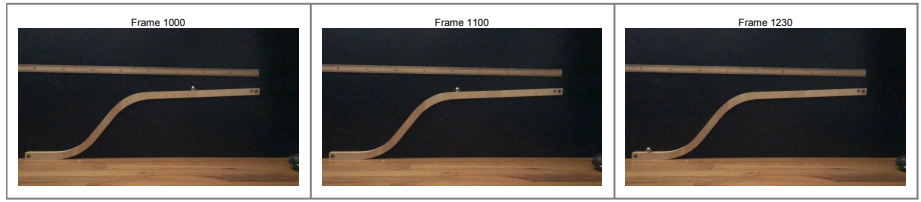}}%
    }
    \caption{Plot shows several frames from a video file. You can notice a small ball moving on the track.}
    \label{fig:ball_on_track}
\end{figure}

\begin{lstlisting}[caption={Automatically extract positions of a ball from a video},captionpos=t]
BallPositions = ImageFeatureTrack[p[[1 ;; 27]], MaxFeatures -> 1]; 
(* takes a long time to compute *)
pPos = ListPlot[BallPositions, ImageSize -> 400, PlotLabel -> "Positions"]
\end{lstlisting}

\begin{lstlisting}[caption={Remove duplicates in x and do interpolation, apply smoothing, interpolate, plot},captionpos=t]
(* Remove records which have the same x *)
(* add frame number first, to keep information about time*)
PositionsWithTime =  MapThread[Flatten /@ Thread[{##}] &, 
                               {BallPositions, Range[1, Length[BallPositions]]}] [[All, 1]];

"Lenght before removal of duplicates is: " <>  ToString[Length[PositionsWithTime]]

(* remove duplicates *)
PositionsWithTimeNoDublicates = Union[PositionsWithTime, SameTest -> (#1[[1]] == #2[[1]] &)]; 
"Lenght after removal of duplicates is: " <>  ToString[Length[PositionsWithTimeNoDublicates]]

(* separate back into time and coordinates *)
time = PositionsWithTimeNoDublicates[[All, 3]];
coord = PositionsWithTimeNoDublicates[[All, {1, 2}]];
pCoord = ListPlot[coord, ImageSize -> 400, PlotLabel -> "Positions, no duplicates"]
\end{lstlisting}

\begin{lstlisting}[caption={Do an interpolation},captionpos=t]
(* if smooting needed 
  smothed = MovingAverage[coord, 2] ;
  ListPlot[smothed]
  ff= Interpolation[smothed];
*)

ff6 = Interpolation[coord];
Minx = Min[coord[[All, 1]]];
Maxx = Max[coord[[All, 1]]];
pFit = Plot[ff6[x], {x, Minx, Maxx}, ImageSize -> 400, PlotLabel -> "Fit"]

\end{lstlisting}

\begin{lstlisting}[caption={Get a derivative},captionpos=t]
(* Derivative *)
pDer = Plot[ff6'[x], {x, Minx, Maxx}, ImageSize -> 400, PlotLabel -> "Derivative"]
Deploy@Grid[{{pPos, pFit, pDer}}, Dividers -> Gray, Spacings -> {2, 2}]
\end{lstlisting}

\begin{figure}[H]
    \centerline{%
        \resizebox{1\textwidth}{!}{\includegraphics{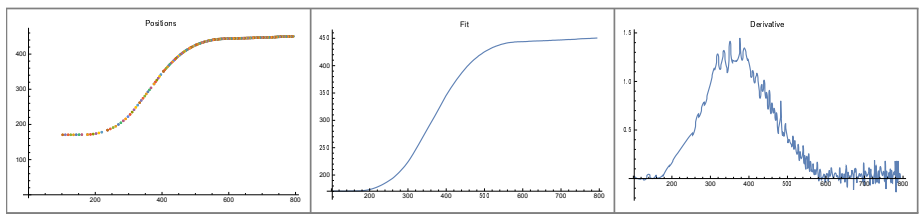}}%
    }
    \label{}
    \caption{Overview of data processing steps. From the left to the right: positions, fit (interpolation), derivative}
\end{figure}

Please note, yet another alternative code for image processing is available for download at this link\footnote{\url{https://github.com/ssamsonau/NumSim_Exp_PhysLabs}}.

\subsection{Doppler shift and Fourier transformation}

\subsubsection{Task with potential subtasks}

Study Doppler effect and use it to produce a sound of an object moving with a given velocity.  Along the way students will get familiar with Fourier transformation and an idea of alternative representations of a signals: using a time domain vs using a frequency domain.

\begin{itemize}
   \item Make a program producing a sound form a moving object given a sound from a standing object
   \item Record sound of a signal of a standing car
   \item Produce sound of moving car using program
   \item Record sound of a signal of a moving car
   \item Compare sound generated numerically with real sound. It is an interesting task for students to think what is the good way to compare predicted signal and recorded signal, as they have to produce an objective way to compare those, including error estimation, etc.
   \item Make corresponding plots
\end{itemize}

\subsubsection{Distribution of work by days} Distribution of a work by 3 days follows the following pattern:

\textbf{Day one.} Concept of Doppler shift and Fourier transformation will be explained, starter code will be given. Some examples with tuning fork or other instruments can be given.  Student will work at home on understanding the underlying theory and adjusting program if needed. 

\textbf{Day two.} An instructor will produce a sound using a horn signal with his/her car, and students will record it with their smartphones. After that the instructor will drive the same car at 25-40 miles/hour near students and produce a horn sound while moving. Students will record this sound as well. Students then have to produce a shifted sound from a "standing recording" using Mathematica and compare it to the recording made for a moving car. 

\textbf{Day three.} Students can use this time as they see fit. 

\subsubsection{Code: Fourier Transformation, starter code} Code for audio data analysis is presented bellow.

\begin{lstlisting}[caption={Example from Mahtematica's help with a note 'G'},captionpos=t]
ClearAll["Global`*"]

record1 = Sound[SoundNote["G", 1, "Violin"]]

Periodogram[record1]  
\end{lstlisting}

\begin{figure}[H]
    \centering
    \begin{subfloat}
        \centering
        \includegraphics[width=.49\textwidth]{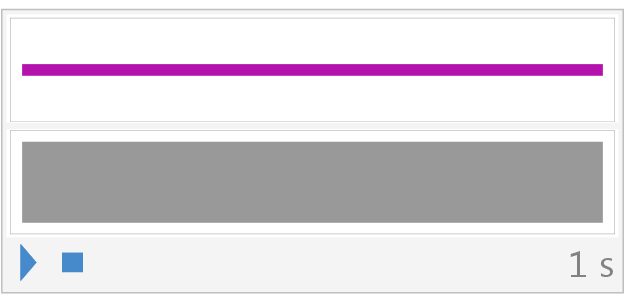}
    \end{subfloat}
    \begin{subfloat}
        \centering
        \includegraphics[width=.49\textwidth]{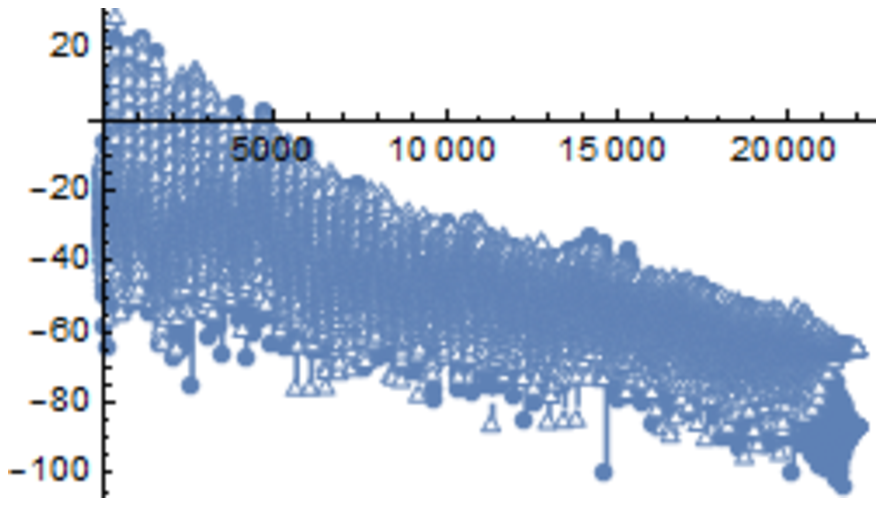}
    \end{subfloat}
    \caption{Sound imported into Mathematica. Left - an interface to play the sound, right - Periodogram}
    \label{fig:sound_periodgram}
\end{figure}

\begin{lstlisting}[caption={Shifting frequencies. Result is shown on fig. \ref{fig:sound_shift}},captionpos=t]

(*shift by adding the same number to all frequencies *)
res=AudioFrequencyShift[record1, Quantity[1000, "Hertz"]]
Periodogram[{record1, res}, PlotRange -> {{0, 4000}, All}, ScalingFunctions -> "Absolute"] 

(* shift by multiplying all the frequencies by the same number *)
new = AudioPitchShift[record1, 1.5]
Periodogram[{new, record1}, PlotRange->{{0,4000}, All}, ScalingFunctions -> "Absolute"]


\end{lstlisting}

\begin{figure}[H]
    \centering
    \begin{subfloat}
        \centering
        \includegraphics[width=.49\linewidth]{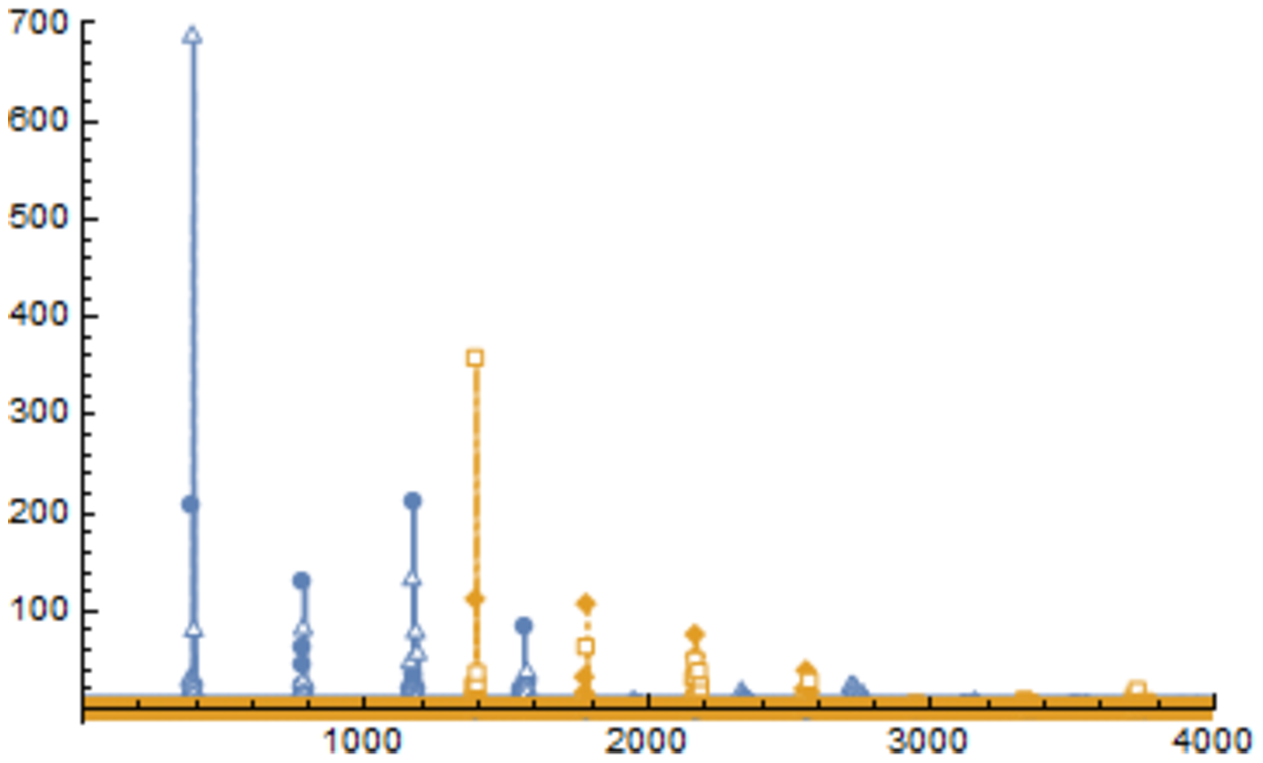}
    \end{subfloat}
    \begin{subfloat}
        \centering
        \includegraphics[width=.49\linewidth]{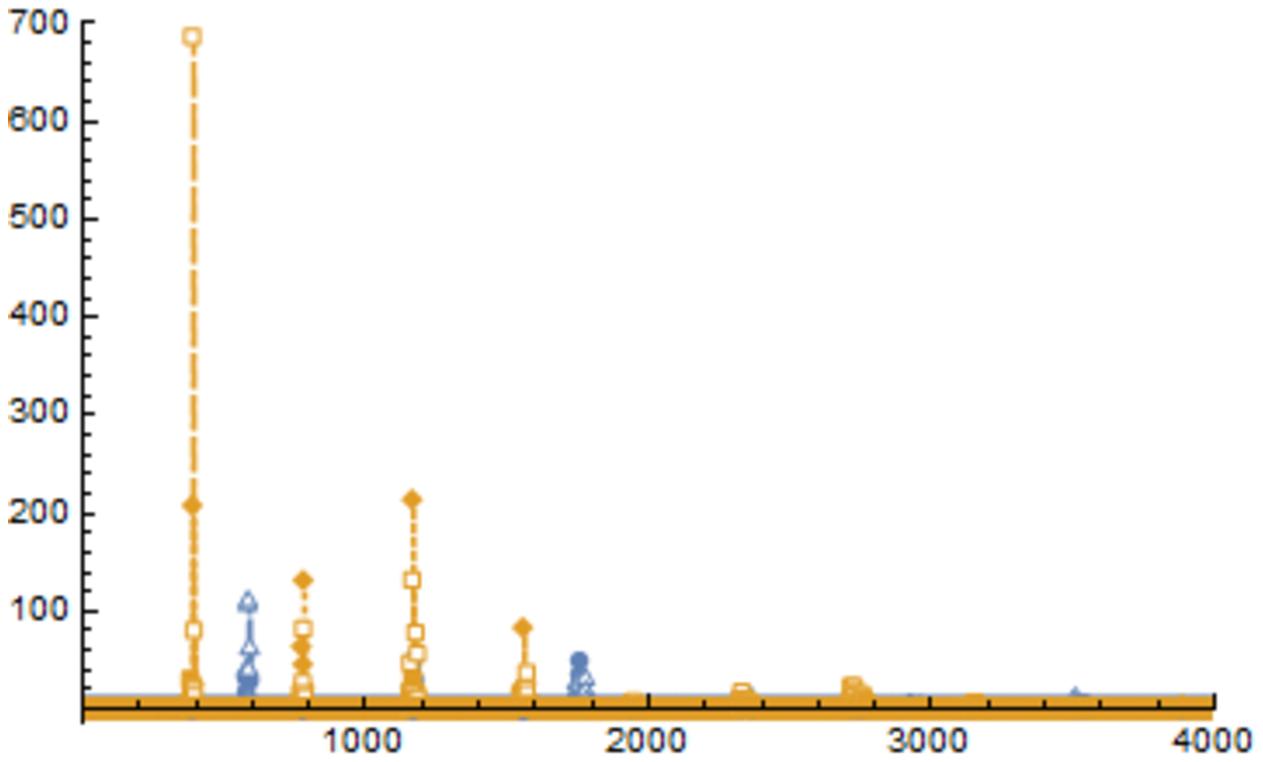}
    \end{subfloat}
    \caption{Periodogram of an original (blue) and shifted (yellow) sound. On the left the shift is done by adding the same number to all frequencies. On the right shift is done by multiplying all frequencies by the same number.}
    \label{fig:sound_shift}
\end{figure}

In order to add an audio recorded by a student, one can just "drag and drop" file into Mathematica.

\begin{lstlisting}[caption={Importing audio file into Mathematica},captionpos=t]
Record2 = (drag and drop audio file)  
\end{lstlisting}

At this stage students have enough knowledge on how to work with sound data to complete the task. Some students may also benefit from functions helping to extract data in a form of a list. Code for such functions is presented bellow.
\begin{lstlisting}[caption={Code for additional operations with audio data},captionpos=t]
(* get Amplitude data in time domain *)
AudioData[new]
AudioData[new][[1]];
AudioData[new][[2]];
ListPlot[AudioData[new][[1]]];
ListPlot[AudioData[new][[2]]];

(* squared magnitude of the discrete Fourier transform (power 
spectrum) of list. *)
PerData = PeriodogramArray[new]
ListLinePlot[PerData[[1]], PlotRange -> {{0, 4000}, All}]

(* automatically find and plot peaks *)
peaks = FindPeaks[PerData[[1]], 100]
ListLinePlot[PerData[[1]], Epilog -> {Red, PointSize[0.02], Point[peaks]},
             PlotRange -> {{0, 4000}, All}
 ]

(* apply filter if needed *)
newFiltered = BandpassFilter[new, {400, 1000}]
PerDataFiltered = PeriodogramArray[newFiltered];
ListLinePlot[PerDataFiltered[[1]], PlotRange -> {{0, 4000}, All}]



\end{lstlisting}

\subsection{Equipotential lines lab}

\subsubsection{Task with potential subtasks}

Use numerical simulations to produce an image of equipotential lines, given the shape of electrodes. Compare to experiment.

\begin{itemize}
   \item Use Mathematica to calculate position of equipotential lines, given interesting (to explore) combination of electrodes in a plane. Note: a relatively simple way to construct irregularly shaped electrodes with Mathematica, is to combine ellipses (see an example in the code)
   \item Print an image produced by Mathematica using a printer
   \item On a conductive paper make the same electrodes arrangement using a conductive ink, or a conductive tape. 
   \item On a printed paper make small holes for a voltmeter probes along the equipotential lines. Align this paper with the conductive paper. Apply voltage to electrodes on the conductive paper. Using a voltmeter measure the voltage in the premade holes. Evaluate how precise numerical predictions are.

\end{itemize}

\subsubsection{Distribution of work by days} 

\textbf{Day one:} starter code and first calculations. \textbf{Day two:} Preparing conductive paper with electrodes. \textbf{Day three:} Final measurements.

\subsubsection{Code: starter code}

\begin{lstlisting}[caption={Define boundaries using mathematical equations. Simple example. The area defined between these boundaries is shown on fig. \ref{fig:area_example} },captionpos=t,
label={lst:boundaries_example}]

bound = {
  x^2 + y^2 - 1 ,
  -x ,
  -y - 0.5
  }
  
(* Apply <= 0 condition to all equations *)
boundConditions = Map[# <= 0 &, bound] 
(* Apply 'and' operation to all conditions *)
boundConditionsAnd = Apply[And, boundConditions] 
(* Region defined by conditions *)
DefinedRegion = ImplicitRegion[boundConditionsAnd , {x, y}] 
RegionPlot[DefinedRegion] (* plot *)

\end{lstlisting}

\begin{figure}[H]
    \centerline{%
        \resizebox{0.3\textwidth}{!}{\includegraphics{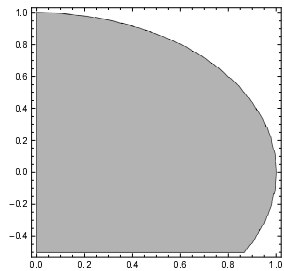}}%
    }
    
    \caption{A gray area is defined as an area between the boundaries specified by equations in listing \ref{lst:boundaries_example}}
    \label{fig:area_example}
\end{figure}

\begin{lstlisting}[caption={Define boundaries using mathematical equations. The area defined between these boundaries is shown on fig. \ref{fig:area_electrical} },captionpos=t,
label={lst:boundaries}]
bound = {
   -y - 0 (*bound[[1]]*), y - 1 (*bound[[2]]*), 0 - x (*bound[[3]]*), -2.5 + x (*bound[[4]]*), 
   1/200 - (-0.2 + x)^2 - (y - 0.5)^2 (*bound[[5]]*), 
   1/200 - (-1 + x)^2 - (y - 0.5)^2 (*bound[[6]]*), 
   1/5 - ((x - 2) Cos[Pi/6] + (y - 0.38) Sin[Pi/6])^2/0.25^2 -
         ((x - 2) Sin[Pi/6] + (y - 0.38) Cos[Pi/6])^2/0.05^2(*bound[[7]]*), 
   1/5 - ((x - 2) Cos[-Pi/6] + (y - 0.62) Sin[-Pi/6])^2/0.25^2 - 
         ((x - 2) Sin[-Pi/6] + (y - 0.62) Cos[-Pi/6])^2/0.05^2(*bound[[8]]*), 
   1/1.25 - ((x - 2.2) Cos[Pi/2] + (y - 0.5) Sin[Pi/2])^2/0.25^2 - 
         ((x - 2.2) Sin[Pi/2] + (y - 0.5) Cos[Pi/2])^2/0.03^2(*bound[[9]]*)};

boundConditions = Map[# <= 0 &, bound] 
boundConditionsAnd = Apply[And, boundConditions] 
DefinedRegion =  ImplicitRegion[boundConditionsAnd, {x, y}]

Show[RegionPlot[DefinedRegion], 
 ContourPlot[
  (* draw color lines on borders within this ranges of x and y *)
  Evaluate[Thread[bound == 0]], {x, -0.1, 5}, {y, -0.1, 2},
  (* colors for lines *)
  ContourStyle -> {Purple, Purple, Purple, Purple, Blue, Red, Green, Green, Green}
  ],
 AspectRatio -> Automatic]

\end{lstlisting}

\begin{figure}[H]
    \centerline{%
        \resizebox{1\textwidth}{!}{\includegraphics{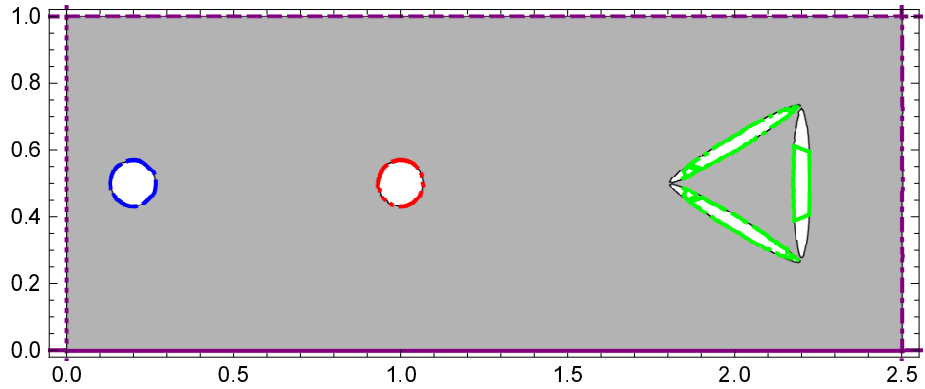}}%
    }
    
    \caption{A gray area is defined as an area between the boundaries specified by equations in listing \ref{lst:boundaries}}
    \label{fig:area_electrical}
\end{figure}

\begin{lstlisting}[caption={Specify voltage on boundaries and solve. Result is shown on fig. \ref{fig:potential}},captionpos=t, label={lst:get_V}]
(*If condition for a boundary not given, potential can have any value there*)

sol=NDSolveValue[{Laplacian[u[x, y], {x, y}]==0,
   {DirichletCondition[u[x, y]==5, bound[[5]]==0],
    DirichletCondition[u[x, y]==0, bound[[6]]==0],
    DirichletCondition[u[x, y]==-2, bound[[7]]==0],
    DirichletCondition[u[x, y]==-2, bound[[8]]==0],
    DirichletCondition[u[x, y]==-2, bound[[9]]==0]}},
    u, {x, y}, Element[{x, y}, DefinedRegion]];

ContourPlot[sol[x, y], Element[{x, y}, DefinedRegion] , 
            Mesh -> None,
            ColorFunction -> "Electric Potential Map",
            Contours -> 18, AspectRatio -> Automatic]

\end{lstlisting}

\begin{figure}[H]
    \centerline{%
        \resizebox{1\textwidth}{!}{\includegraphics{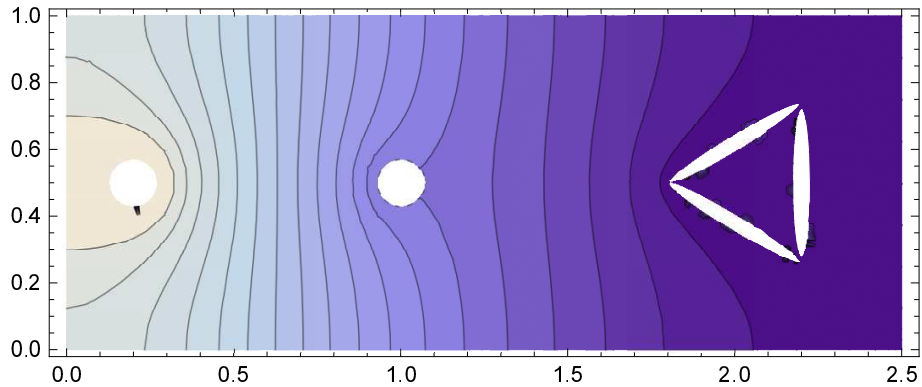}}%
    }
    \caption{Electric potential contour plot calculated using code in listing \ref{lst:get_V}}
    \label{fig:potential}
\end{figure}

\section{Conclusion}
Above we present the complete set of 7 laboratory modules ready to be implemented in advanced high schools or colleges. Of course, this sequence is not set in stone and an adjustment can/should be made by an instructor to adapt for a particular group of students. This is our hope, that this paper will serve as a basis for a development of many other laboratory courses combining experiments with computer simulations. It would be interesting to perform a study involving many students to obtain statistical data on pedagogical outcomes.

\section{Acknowledgements}
We express gratitude to students of Princeton International School of Mathematics and Science for their effort, open minded approach, and interesting ideas about implementation of some of the topics. We appreciate the support of administration and faculty of Princeton International School of Mathematics and Science.

\bibliographystyle{unsrt}
\bibliography{main}

\end{document}